\documentclass[a4paper,fleqn,usenatbib]{mnras}

\usepackage{newtxtext,newtxmath}

\usepackage[T1]{fontenc}
\usepackage{ae,aecompl}

\usepackage{graphicx}	
\usepackage{amsmath}	

\newcommand{\HST}{{\it HST}}
\newcommand{\IUE}{{\it IUE}}
\newcommand{\HeI}{\ion{He}{I}}
\newcommand{\HI}{\ion{H}{I}}
\newcommand{\NII}{\ion{N}{II}}
\newcommand{\OI}{\ion{O}{I}}
\newcommand{\OIII}{\ion{O}{III}}
\newcommand{\SII}{\ion{S}{II}}

\title[Stingray Nebula and V839 Ara]{Startlingly Fast Evolution of the Stingray Planetary Nebula and its Central Star, V839 Arae\thanks{Based on observations with the NASA/ESA {\it Hubble Space Telescope\/} obtained at Space Telescope Science Institute, operated by Association of Universities for Research in Astronomy, Inc., under NASA contract NAS5-26555.}}

\author[B. E. Schaefer et al.]{
Bradley E. Schaefer$^{1}$\thanks{E-mail: schaefer@lsu.edu},
Howard E. Bond$^{2,3}$,
Kailash C. Sahu$^{3}$
\\
$^{1}$Department of Physics and Astronomy, Louisiana State University, Baton Rouge, LA 70820, USA\\
$^{2}$Department of Astronomy and Astrophysics, Pennsylvania State University, University Park, PA 16802, USA\\
$^{3}$Space Telescope Science Institute, 3700 San Martin Drive, Baltimore, MD 21218, USA\\
}

\date{Accepted XXX. Received YYY; in original form ZZZ}

\pubyear{2020}

\begin{document}
\label{firstpage}
\pagerange{\pageref{firstpage}--\pageref{lastpage}}
\maketitle

\begin{abstract}

The planetary nebula (PN) called the Stingray (PN G331.3$-$12.1) suddenly turned on in the 1980s, and its central star (V839 Ara) started a fast evolution with large amplitudes in magnitude, surface temperature, and surface gravity, perhaps as part of a late thermal pulse causing a loop in the Hertzsprung-Russell (HR) diagram.  With these fast changes, we have taken images with the {\it Hubble Space Telescope} in 2016.  We do {\it not} see the massive high-velocity mass loss of the 1980s, either close in to the central star, or as clumps or changes in the nebula far from the star, or as localized or general increases in the emission-line flux caused by the shocks of the outflowing ejecta ramming into the slow-moving PN shell.   We think that the lack of seeing the outgoing material is because the relatively modest total mass had already suffused the PN before the first resolved imaging in 1992, and it was the shocks from this impact that initially ionized the Stingray starting in the 1980s.  We also quantify the complex fast fading of the Stingray, with each emission line and each structure having different fade rates, with half-lives ranging from 3 to 29 years.  In a century or two, the PN will fade to invisibility.  With this complex fading of different structures, it is impossible to derive any expansion rate for the PN\null.  The central star had its brightness roughly constant from 1996 to 2016, but with substantial erratic variability from 15.50 to 14.18 mag in the $V$ band. 

\end{abstract}




\section{Introduction}

The ``Stingray''\footnote{The ``Stingray'' name was proposed by the young children of M.~Bobrowsky, when shown the initial {\it Hubble Space Telescope\/} images.} planetary nebula (PN) has shown extraordinarily rapid evolutionary changes over the past several decades. Its central star (CPD~$-$59$\degr$6926, Hen 3-1357, SAO 244567, V839~Ara) was catalogued as an unremarkable H $\alpha$-emission star by Henize (1976). But in the 1980s it suddenly developed strong emission lines characteristic of a young PN (now catalogued as PN~G331.3$-$12.1), and the central star began rapid evolution in brightness and effective temperature. We appear to be witnessing unique and surprisingly rapid post-asymptotic-giant branch (post-AGB) stellar evolution, resulting from a late thermal pulse (LTP).

Before about 1980, all available spectra showed either weak Balmer emission lines or no emission lines (Schaefer \& Edwards 2015).  During this time, the spectrum was that of a normal B0 or B1 star (Parthasarathy et al.\ 1995), placed into luminosity class I or II\null.  The star slowly began to attract attention, first for having just some H $\alpha$ emission (Henize 1976), then as an {\it IRAS\/} far-infrared source selected as a proto-PN (Parthasarathy \& Pottasch 1989; Volk \& Kwok 1989).  By 1990 and 1992, the optical spectrum of the star was dominated by very bright and narrow [\ion{O}{III}] emission lines, plus other lines that are those of a young PN (Parthasarathy et al.\ 1993, 1995).    The stark difference between the 1979.49 spectrum (showing an O/B star with no emission lines) and the 1988 {\it IUE} spectrum (very bright PN emission lines) shows that the Stingray is evolving rapidly, with the PN having ``turned on'' starting around 1980 when the pre-existing neutral shell was ionized.

This ionization event in the 1980s happened in stark coincidence with a mass loss event (a mass ejection event) in the 1980s.  \IUE\/ observed the velocity of a fast wind during the 1980s event to be 1800 km s$^{-1}$ or larger as based on the P Cygni profiles, with a high mass-loss rate (Feibelman 1995).  This mass loss was presumably not present before 1980, was not quantitatively measured before 1988, was close to 10$^{-9}$ M$_{\odot}$/year from 1988-1993, then dropped by 10$\times$ in 1994, and has kept dropping by orders of magnitude at least up until 2006 (Reindl et al.\ 2014).  So we have a high-velocity mass loss event confined in time to roughly 1980 to 1993.  The velocity and rate are comparable to that of stellar winds for such stars (Cerruti-Sola \& Perinotto 1985; Pauldrach et al. 1988).  But the Stingray's mass loss mechanism cannot be anything we would ordinarily call as a stellar wind because stellar wind mechanisms all produce variability only on very long time scales (Villaver et al. 2002, Perinotto et al. 2004), while observed variability in PN central stars is always small than a few per cent (Guerrrero \& De Marco 2013).  So we do not have any useable idea as to the cause and mechanism for the acceleration of the mass loss.  Nevertheless, the stark coincidence in time between the 1980s ionization event and the 1980s mass loss event carries a strong case that the two are causally related.

A compact PN shell was resolved in 1992 by the {\it Hubble Space Telescope\/} ({\it HST}) with a bipolar shape and an embedded `equatorial ring', named the `Bright Inner Ring', with an inclination of 56$\degr$ (Bobrowsky 1994; Bobrowsky et al.\ 1998).  Reindl et al.\ (2014) present an analysis in which they derived a kinematic age of the observed PN shell to be 1013$^{+488}_{-793}$~yr, based on an angular radius of 1.15 arc-seconds, a derived spectroscopic distance of 1.6$^{+0.8}_{-1.2}$ kpc, and an expansion velocity of 8.4 km s$^{-1}$ (based on the [\ion{O}{III}] 5007 \AA\ line width).  This shows that the observed PN shell is many centuries old, and was ejected long before the 1980s ionization event made the nebula optically visible.

The central star faded slowly from $B=10.30$ in 1889 to 10.76 in 1980, in a surprising anticipation of the 1980s ionization event (Schaefer \& Edwards 2015).  Also surprisingly, the 1980s ionization event did {\it not\/} produce any brightening in the $B$~band (Schaefer \& Edwards 2015).  For yet another startling phenomenon in the photometry of the central star, from 1980 to 1997, the brightness rapidly faded, at a rate of 0.20 mag~$\rm yr^{-1}$, dimming to $B=14.64$ in 1997 (Schaefer \& Edwards 2015).  The central star was resolved from the nebula in 1996 March with \HST, and was reported at $V=15.4$, as deduced from the flux measured in a continuum filter centred at 6193~\AA\ (Bobrowsky et al.\ 1998).  Reindl et al.\ (2014, 2017) report the temperature and surface gravity of the central star, mapping out the surprisingly fast changes.  They concluded that the best idea is that V839 Ara is undergoing a LTP.   An LTP is an event in the lifetime of a post-AGB star, passing right-to-left across the top of the HR diagram, still with an outer hydrogen envelope, where convection triggers helium shell burning that last for some number of centuries (Bl\"{o}cker 2001, Sch\"{o}nberner 2008).

V839 Ara and the Stingray Nebula are fast changing.  So it is worthwhile to follow the on-going evolution, and {\it HST\/} is the only means to get updates, with the last {\it HST\/} images were in 2000.  This paper reports on our observations with {\it HST\/} in 2016.  Our goals were to seek the fast-moving massive material from the 1980s ionization event, to measure the expansion of the PN, and to update the stellar photometry to follow it through the LTP loop.

\section{Observations}

We were granted two orbits of time with the WFC3/UVIS camera on the {\it HST\/} on 2016 February 21.  We took long and short exposures through filters F438W ($B$) for 4 and 120~s, F555W ($V$) for 4 and 60~s, and F625W ($r'$) for 4 and 90~s.  Further, we took long exposures through five narrow-band filters: F487N (H $\beta$) for 600~s, F502N ([\ion{O}{III}]) for 600~s, F656N (H $\alpha$) for 382~s, F658N ([\ion{N}{II}]) for 600~s, and F673N ([\ion{S}{ii}]) for 360~s.  

For our science, we need an early and late epoch of comparable images.  Our 2016 {\it HST\/} images provide the late epoch.  For the broad-band filters, the only early epoch wide-band {\it HST\/} images are a pair of F555W images (30 and 180~s) from 2000 March 2 with WFPC2.  For narrow-band filters, available early-epoch images are from 1992 August 12-23 (PI=M.~Bobrowsky) with WFPC1, 1994 March 8 (PI=M.~Bobrowsky) with WFPC2, and 2000 March~2 (PI=A.~Hajian) with WFPC2.  The 1992 images are only in the F502N and F487N filters, and they suffer from the optical aberration of the just-launched {\it HST}; see Bobrowsky (1994).  The 1996 images were all through narrow-band filters, including the five we used again in 2016, plus F588N (\ion{He}{I}) and F631N ([\ion{O}{I}]); see Bobrowsky et al.\ (1998).  The 2000 narrow-band images were only with filters F502N ([\ion{O}{III}]) and F658N ([\ion{N}{II}]).  

All of these {\it HST\/} images are publicly available through the {\it Mikulski Archive for Space Telescopes}.  Data processing was performed with the usual {\it HST\/} pipeline.  Our analysis of the {\it HST\/} images was entirely with standard {\it HST\/} calibrations and IRAF routines.

Fig. 1 shows a montage of the available narrow-band images from 1996, 2000, and 2016.  Fig. 2 shows a false-colour combination of the H $\beta$ (for the blue light), [\ion{O}{III}] (for the green light), and [\ion{N}{II}] for the red light, both for the 1996 and 2016 data sets.

\section{No Emerging Shell}

In the 1980s, V839 Ara ejected a fast wind with terminal velocity of 1800 km s$^{-1}$ or faster (Feibelman 1995; Reindl et al.\ 2014).  For this velocity and a distance of 1.6 kpc, the undecelerated wind will reach a radius of 1 arc-second in 15 years.  The observed radius of the Stingray nebula is roughly 1 arc-second, so by 15 years after ~1985 (i.e., anytime after the year 2000), the fast ejecta could have reached everywhere in the PN shell.  With possible decelerations, the inevitable slower moving material, and projection effects, the 1980's ejecta might appear partly as clumps near the center and around the central star.

The massive wind started around 1980, but during the active ejection phase of the 1980s, the only available measures of the mass-loss rate are 3$\times$10$^{-8}$ M$_{\odot}$/year for 1988 (Feibelman 1995) and close to 10$^{-9}$ M$_{\odot}$/year in 1988, 1992, and 1993 (Reindl et al.\ 2014).  By 1994, the star's mass-loss rate had tapered off to 10$^{-10}$ M$_{\odot}$/year (Reindl et al.\ 2014).  Based on this sketchy information and adopting the rates from Reindl et al., for roughly a decade of ejection, we can estimate that the total mass ejected was $\sim$3$\times$10$^{-7}$ M$_{\odot}$.  

This situation resembles the case for recurrent nova T~Pyx, where the later outgoing shells sculpted and lit up the outer shells (Schaefer et al.\ 2013; Shara et al.\ 2015).  Another similar case of a fast inner ejecta shell lighting up an outer PN shell, and later shock-heating the shell, is that of SN~1987A (Gr\"{o}ningsson et al.\ 2008), albeit with higher mass and velocity than for V839 Ara.  With the observed fast and massive ejecta from the central star, we should be seeing the Stingray lighting up.

The outgoing shell of massive and fast-moving material from the 1980s should be visible in our images in all of five ways:

\subsection{Images of the Inner Region}

The outgoing material will be highly ionized, and should be brightly emitting light, primarily in the usual emission lines prominent in PNe.  This material should appear in our narrow-band imaging, mainly close in to the central star.  This is the same task as for seeking nova shells close-in around recent novae.  The most direct detection method is simply to look for nebulosity around the central star in direct images through narrow-band filters.

In Fig. 3, we display a 60~s exposure through the F502N ([\ion{O}{III}]) filter.  Most of the region in the interior of the Bright Inner Ring shows a flat fill with no features.  Indeed, it is this lack of features that shows that no outgoing shell is visible.  The lack of any visible outgoing shell is also apparent in all our other images, both broad-band and narrow-band.

\subsection{Radial Profiles Around the Central Star}

It is possible that the light from the outflowing 1980s ejecta is primarily within the point-spread function (PSF) of the central star.  Our second method to recognize the ejecta is to look at the radial profile of the central star, where deviations from the PSF could point to extra light from the inner shell.  For example, a shell of ejecta with an angular radius of $\sim$0.1 arc-seconds might be apparent as a tail in the radial profile, or as an increased effective radius (measured as a Full-Width-Half-Maximum; FWHM) of the central star image.  As we strongly expect the ejecta to be emitting most of its optical light in the usual PN emission lines, and the central star itself will be faint in narrow-band filters, our most sensitive search will be in our narrow-band images.

We have constructed radial profiles with IRAF, and this does a good job at centroiding the stellar image, identifying the background, and fitting a Gaussian PSF\null.  Here, we can take the average and  root-mean-square (RMS) of the three IRAF measures of FWHM as a quantification of the stellar size and its uncertainty.  We take the radial profiles of the companion star and of nearby stars outside the PN to provide the PSF of the images.  To an accuracy that is substantially better than the observed scatter in the radial profiles, the star images are well fit to the Gaussian profiles.  In particular, the radial profile of V839 Ara is closely fit with a Gaussian, with no significant deviations, in all our images.  Further, the central star has a FWHM identical to that of the other stars.  For example, for the 600~s F502N image, the FWHM of the central star is 2.09$\pm$0.09 pixels, while the FWHM of a nearby star is 2.13$\pm$0.05 pixels.  For an example from a broad-band filter, the 60~s image in F555W has the central star with a FWHM of 2.08$\pm$0.08 pixels, while the comparison star has a FWHM of 2.07$\pm$0.06 pixels and a nearby star has a FWHM of 2.15$\pm$0.04 pixels.  All this goes to show that any outgoing ejecta are not hiding in the tail of the central star's PSF.

\subsection{Subtraction of the Central Star}

Our third method to seek the outflowing ejecta is to subtract the PSF from the central star, and see what remains.  If the shell is itself nearly Gaussian shaped with a radius similar, but different to the {\it HST\/} PSF, then this method will subtract out the emerging inner shell.  This third method is good for catching most shells that we would be familiar with, and in particular, those that are asymmetric or even jet-like.  The trick here is to get the PSF to subtract out.  We could use the standard {\it HST\/} tool for deriving the PSF, but this is very sensitive to input, does not account for effects like the telescope's slight changes in focal length (i.e., breathing), and does not have the highest accuracy.  For our 2016 images, we cannot just use the stars in the field, as these have different positions in the field of view (and hence slightly different PSFs), and they are all fainter than V839 Ara so the statistics will be poor.  So we have adopted the following method, based on the very good presumption that any emerging shell will be ionized and be mostly an emission line spectra, with H $\alpha$ and [\ion{O}{III}] dominating.  The idea is to use the F555W ($V$-band) image to provide the PSF for subtracting from the F502N ([\ion{O}{III}]) image.  This has a big advantage that the PSFs for the two images are taken within a few minutes and the wavelengths are a good match and the position on the chip is a match, so this should really be a good PSF\null.  Any shell light in the $V$-band image will be a very small fraction of the continuum light, so we would only be subtracting off $<$1 per cent of the shell light, and this is negligible.  So this is a very good way to get the PSF of the central star for subtracting off the stellar light, leaving virtually all the shell emission-line light.  Also, we can use the F625W ($r'$) image to form the PSF for the continuum light from the central star in the H $\alpha$ images.  

The broad-band images must first be shifted and scaled before being subtracted out from the narrow-band images.  In practice, we chose the shifts and the scale factors so as to make the subtracted image the flattest possible in the central region.  If the shell is so small as to be well below the {\it HST\/} resolution, then its light will be spread out with the PSF, so then there is no way to use imaging to distinguish the shell.  But if the shell is even marginally resolved, then its extra flux must have a FWHM larger than the PSF, so there must still be extra flux visible after the PSF subtraction.  This third method is substantially more sensitive than the second method for asymmetric ejections, like for jets.

We have performed these two PSF subtractions, with the PSF-subtracted F656N (H $\alpha$) image displayed in Fig.~4.  We see the complete removal of any deviation from a flat background inside the Inner Bright Ring.  That is, there is no visible flux from any ejected shell hiding under the central star.  Similarly, the PSF-subtracted F505N ([\ion{O}{III}]) image shows no significant flux in the centre of the Bright Inner Ring.  That is, we again can place severe limits on any emission lines from the 1980s ejecta.

To be quantitative, we can constrain the flux in the 1980s ejecta as a fraction of the total line emission of the Stingray.  For the PSF-subtracted H $\alpha$ image (see Fig. 4), the extra flux above a flat background is $<$1 ADU in any single pixel and $<$3 ADU in any 3$\times$3 pixel square, while the entire nebula is 66,500 ADU, so the fractional H $\alpha$ flux in the ejecta's shell is less than 45 parts per million.  For the PSF-subtracted F502N ([\ion{O}{III}]) image, the extra flux above a flat background is $<$5 ADU in any single pixel and $<$2 ADU in any 3$\times$3 pixel square, while the entire nebula is 62,000 ADU, so the fractional [\ion{O}{III}] flux in the ejecta's shell is less than 32 parts per million.  Note that these limits apply only for the emission from a part of the outgoing shell close-in to the central star.

\subsection{Unresolved Brightness Change in the $V$ band}

The shocks created by the 1980s ejecta colliding with the inner regions of the PN shell will necessarily heat and ionize the gas, making for extra light in the usual PN emission lines.  That is, when the shocks start up, perhaps sometime around the year 2000, the [\ion{O}{III}] line flux must rise above the `background' caused by the 1980s ionization event.  It is difficult to make absolute calibrations of spectrophotometry that are accurate over the decades and over many instruments.  

Fortunately, the $V$ bandpass is completely dominated by light from the [\ion{O}{III}] lines (e.g., fig. 2 of Parthasarathy et al. 1993), so a well-calibrated $V$-band magnitude is a direct measure of the total [\ion{O}{III}] flux from the Stingray Nebula.  That is, the total $V$ magnitude from unresolved ground-based observations give well-calibrated magnitudes that are essentially just [\ion{O}{III}] fluxes.  Getting consistent $V$-band photometry across the years is also difficult to do accurately, because the various observers will be using detectors with slightly varying spectral responses.  Fortunately, we do have several surveys that provide long-term $V$-band light curves with consistent instrumentation (see Schaefer \& Edwards 2015).  These include CCD photometry from the ASAS project, providing 425 measurements from 2001--2009, and the visual photometry of Albert Jones (arguably the best variable star observer of the last century, renowned for his photometric accuracy), with 128 magnitudes from 1994--2007. Further, the APASS CCD light curve in the $V$ band provides 352 measurements from 2011--2015.  Full observational details, analysis, and light curves for ASAS, Jones, and APASS are given in Schaefer \& Edwards (2015), see their fig. 2.  All of these measures are essentially recordings of the [\ion{O}{III}] flux from the entire nebula.  They reveal that the $V$ magnitude has been steadily fading at a rate of roughly 0.1 mag/year from 1994--2015 (Schaefer \& Edwards 2015).  Schaefer \& Edwards (2015) also chronicle the fading of the nebular light in the near-infrared, middle-infrared, and radio. This fading is due to the ordinary recombination inside the nebular gases after the 1980s ionization event, where the various physical emission mechanisms are governed by the recombination, yielding the observed fading time-scales (Schaefer \& Edwards 2015).

Thus the $V$-band light curve is a good measure of the overall [\ion{O}{III}] flux from the Stingray Nebula from 1994--2015 (see fig. 2 of Schaefer \& Edwards), and the collision of the 1980s ejecta with the inner edge of the PN starting after the 1990s should cause a brightening of the [\ion{O}{III}] flux. However, such a brightening is certainly not seen in the light curve.  So on the face of it, the outgoing ejecta are invisible.  A better statement is that the flux increase due to the shocks is actually less than about 10 per cent of the total [\ion{O}{III}] flux for a time somewhat from 1994 to 2015.  Still, as the impact speed is 1800 km s$^{-1}$ and should create large amounts of flux, it is surprising that no brightness flare can be associated with the ramming of the ejecta into the PN.

\subsection{Surface-Brightness Changes 1996--2016}

Another way to seek the outgoing 1980s ejecta is to use the resolved {\it HST\/} images to look for regions or structures in the nebula that have brightened recently.  This brightening should be seen in the narrow emission lines, and the baseline from 1996 to 2016 should show the brightening.  To perform this test, we made three analyses with increasing quantification.  The first analysis is to simply visually compare the 1996 and 2016 images, looking for any region or structure that brightened relative to other structures.  None was found, which is to say that there are no obvious brightening (from the shock ramming into existing structures) nor any apparent new structures (from the outgoing material).  The second analysis was to construct brightness profiles running through the central star, with these being discussed in detail in Section 5, for which we see no evidence of brightening structures or new structures.  The third analysis was to quantitatively calculate the 1996--2016 fade-ratios ($\mathcal{F}$) for many points inside the PN, as described in Section 4.2.  For this, we see the various parts of the nebula fading at different rates (correlated with the nebular brightness at each point), but all points in the nebula are {\it fading\/}, not brightening.  

In all, by five different methods, we are seeing neither the outgoing 1980s ejecta nor the shocks of it ramming into the pre-existing PN.

\section{Structures of the Stingray}

The {\it HST\/} images of the Stingray Nebula show a complex structure with many arcs of light (see our Fig. 1 to 4).  See fig. 2 of Bobrowsky et al.\ (1998) for the names of various structures.  The brightness levels within the images are proportional to the column density of ionized gas, so the images directly show the relative mass distribution of the shells and rings.  

Various additional information can be gleaned about the structure of the Stingray Nebula by constructing ratios of images in different emission lines.  That is, with shifting to co-align the images, then a ratio of the images will have each pixel proportional to the local ratio of the line fluxes.  In constructing our ratio images, with the background light always being near zero, we have set a floor of 1.0 ADU in the input images, as that way the background regions will appear uniform (rather than very noisy).  The intrinsic width of the Stingray lines is greatly smaller than the width of the {\it HST\/} filters, so all the line flux is included.  The central transmissivity of all the filters we used is similar (near 25 per cent), and the WFC3/UVIS quantum efficiency is fairly flat across the optical range, so the observed ratios are close to the actual photon flux ratios for that part of the nebula.

\subsection{Ionization Levels}

A primary usage for our ratio images is to map out the relative ionization levels throughout the nebula.  That is, the ratio images will show the relative brightness of two lines, all on a pixel-by-pixel basis, where the individual ratios are a function of the ionization level of the gases.  These ratio images should not be directly dependent on the nebular brightness, so structures seen in the ratio images should reveal primarily the ionization.  Line ratios are a complex function of the ionization level, the gas density, the local composition, and various atomic constants.  So it is generally difficult to pull out detailed physical conditions from our line ratios, yet the ratio images can still display large qualitative differences.  For reference, the ionization potential for \ion{H}{I} is 13.6 eV, for \ion{He}{I} is 24.6 eV, for \ion{N}{II} is 29.6 eV, for \ion{O}{I} is 13.7 eV, for \ion{O}{III} is 54.9 eV, and for \ion{S}{II} is 16.3 eV.  Thus the H $\alpha$, H $\beta$, [\ion{O}{I}], and [\ion{S}{II}] lines have similar ionization potentials, which are substantially lower than those of the other lines, while [\ion{O}{III}] has by far the highest ionization potential.
	
What we see consistently is that the ratios are fairly uniform across the nebula, except that the outermost regions are extreme.  

The H $\beta$/[\OIII] ratio is greatly higher in the outermost edge, mainly the edge of the Outer Shell, but also along the outer edge of the Outer and Inner Collimated Flows.  This ratio image is displayed in Fig. 5.  We see the identical situation for the H $\alpha$ and [N II] images divided by the [O III] image.  This means that the outermost edges have H $\beta$ flux relatively much larger than [O III] flux.  The oxygen line has a greatly higher ionization potential than the H $\beta$ line.  We take this as a case of well-known ionization stratification in PNs (Osterbrock 1964).  That is, the spectral lines with the highest ionization potential are emitted in regions close to the central star, while spectral lines with lower ionization potential are relatively more prominent around the edges of the PN. 

We have looked back at the [\OIII] and [\OI] images from 1996.  The [\OI] data were noisy, despite the 1100 seconds of exposure.  The [\OIII]/[\OI] ratio should be an indicator of the ionization level.  We see a ring of low ionization all around the outside edge, tracing the entire edge of the Outer Shell plus the outer edge of the Outer Collimated Outflow (but only over the lobe to the northwest).  The highest ionization is along the Bright Inner Ring.  This is another example of the ionization stratification.  This is in 1996, when we might not have any shock from the 1980s ejecta.

The H $\alpha$/[\SII] ratio image has different structure.  There is relatively little H $\alpha$ emission around the outer edges, with most coming from around the Bright Inner Ring.  We attribute the different structure (as compared to the [\OIII] lines) to both \HI\ and [\SII] having similar ionization potentials, so that they are both ionized to equal levels, at least originally.  But even this, where the original ionization levels should be the same, is hard to interpret as the structure in the ratio image will be a complex function with strong time-dependencies of the density, temperature, abundances, and initial degree of ionization.

\subsection{Secular Fading}

Since 1990, the $V$-band light curve has been almost entirely from [\OIII] emission-line flux, so the ground-based light curves of the unresolved system is a good measure of this one line flux.  The $V$-band light curve from 1994 to 2018 shows a steady decline, with the total fading from 1994 to 2015 being by nearly a factor of 6 (Schaefer \& Edwards 2015).  So we should see this factor of six decline in comparing the F502N images from 1996 and 2016, except with our {\it HST\/} images we can measure this decline on a pixel-by-pixel basis.  Thus, we can measure the decline rates for each structure.  We expect variations due to the effective densities being different between structures, where the higher-density regions will recombine faster and hence fade faster.

In comparing the surface brightness of the 1996 and 2016 F502N images, we cannot simply compare the values in the pixels as quoted in the images produced by the {\it HST\/} pipeline.  The images have different pixel sizes, exposures, detector quantum efficiencies, filter transmissions, and detector gains.  The pipeline has already divided by the exposure times, and we can scale by the pixel area so as to get surface brightnesses.  For dividing pixel values between epochs, most of the remainder of the corrections are lumped into the parameter PHOTFLAM, which is tabulated in each header.  We have made an empirical cross calibration based on constant field stars outside the Stingray Nebula, and their total fluxes in 1996 and 2016 are closely given by the correction factors we get from the exposure times and PHOTFLAM\null.  To give an example for comparing surface brightnesses at points in the nebula, the pixel values (fluxes in units of `counts/sec') for the 1996 F502N image can be converted to the same scale for the 2016 F502N image by multiplying by the ratio of exposures (600 s divided by 120 s, or a correction factor of 5.0), by multiplying by the ratio of the pixel areas ([0.03962"/0.10000"]$^2$, or a correction factor of 0.1570), and by multiplying by the ratio of PHOTFLAM values (a correction factor of 59.01).  The pipeline-processed images are already corrected for the exposure, so the 1996 image needs to be multiplied by a factor of 0.1570$\times$59.01 = 9.26, so as to get the values for both the 1996 and 2016 images to be on a consistent scale proportional to surface brightness.

With this, we have tabulated the fade-factor, $\mathcal{F}$, as the ratio of the earlier-and-brighter surface brightness to the 2016 surface brightness in the same filter and for the same point in the nebula.  These are given in Table 1.  Again, the terminology for the structural parts of the Stingray Nebula are given in fig. 2 of Bobrowsky et al.\ (1998).  The $\mathcal{F}$ values are given for 1996--2016 and 2000--2016 in the F502N ([\OIII]) filter, as well as for 1996--2016 in the F656N (H $\alpha$) filter.  Also given are the flux values (in units of electrons per second inside the pixel) for the 2016 images.

We see that the rates of fading vary widely within the nebula, ranging over 4.6--99 in the [\OIII] line and ranging over 1.0--2.3 in the H $\alpha$ line.  The 1996--2016 $\mathcal{F}$ values are tightly correlated with the 2000--2016 $\mathcal{F}$ values, with the ratio of the fade rates equal to the durations of the two time intervals, so it appears that the fade rate is constant.  The 1996-2016 $\mathcal{F}$ values in [\OIII] are reasonably correlated with the 1996-2016 $\mathcal{F}$ values in H $\alpha$, although the two points in the halo are far outliers.

Outside of the Bright Inner Ring, mainly towards the northeast and towards the southwest (out along the major axis of the Bright Inner Ring), inside the Outer Shell, a faint halo of light appears.  This Outer Shell light behaves differently from the much brighter light making up the bulk of the nebular light.  From 1996--2016, the [\OIII] line flux faded very fast, roughly one order of magnitude faster than everywhere else in the Stingray.  Over the same time interval, the H $\alpha$ line flux in this part of the Outer Shell faded at a rate comparable to the rate for other parts of the PN\null.  We do not understand the reason for the very high $\mathcal{F}$ in [\OIII] light only.  But apparently, the Outer Shell has a different gas and radiation dynamics than the rest of the nebula.

The $\mathcal{F}$ values are strongly correlated with the [\OIII] flux, $F$, with an exponential relation.  Plotted on a  $\ln(F)$ versus $\mathcal{F}$ graph, the Table 1 points for [\OIII] form a straight line, which means that the surface brightness is proportional to $e^{-0.31\mathcal{F}}$.  This relation ignores the two faint points in the Outer Shell that are far outliers to this correlation.  However, for H $\alpha$, the situation is not so simple, because the six faintest points (all in the outermost collimated outflows or the Outer Shell) do not have the highest fade rates.  Taking all points in Table 1, the H $\alpha$ surface brightness is proportional to $e^{-0.72\mathcal{F}}$.  Excluding the six faintest points (the last six lines in Table 1), the surface brightness is proportional to $e^{-1.1\mathcal{F}}$.

PN with a fast fading central star are expected to have a fading nebulosity, as the shell recombines.  Such conditions are uncommon, although the fading should be observable for He 1-5 (Arkhipova, Esipov, \& Ikonnikova 2009, Jurcsik \& Montesinos 1999) and HuBi 1 (Guerrero et al. 2018).  To the best of our knowledge, no PN has ever been seen to have its surface brightness fading over the years.  So the Stingray provides a good laboratory for the gas physics.  The obvious idea is that fading arises because the nebula is rapidly recombining, so the falling rate of line emission is just simply due to the fast-falling density of the ions and the free electrons.  This speedy recombination would account for the fast fading of the nebulosity we see in the various optical emission lines, as well as seen in the near-infrared, far-infrared, and radio (Schaefer \& Edwards 2015).

The recombination rates are proportional to the ion density ($N_{ion}$) times the free electron density ($N_e$) times atomic constants.  (These atomic constants will have dependencies on the gas temperature, but Otsuka et al.\ (2017) show that the gas temperature is apparently nearly constant.)   $N_e$ changes over time, being dominated by the recombination of the hydrogen.  The 1996-2016 fade rate will be $\mathcal{F} = [N_{ion}N_e]_{1996}/[N_{ion}N_e]_{2016}$.  Different ions, for example O$^{2+}$ and H$^+$, will have greatly different recombination rates, so their $N_{ion}$ values will change differently.  Thus the 1996-2016 $\mathcal{F}$ values will vary from emission line to emission line.

The fade rates reported in Table 1 show that the brightest parts of the nebula are the slowest at fading.  The brightness along the line of sight scales as the peak gas density around the position of highest gas concentration.  The recombination rate  (and hence the emission line brightness) will be proportional to the $N_{ion}$ and $N_e$.  And the higher the recombination rate, the higher the fading rate.  So the brightest parts of the nebula should have the fastest rate of fading.  That is, the bright regions are the densest regions, hence recombining at a fast rate and using up the free electrons speedily.  But this is the opposite of what we see.  We do not know the resolution of this paradox.

The emission-line flux from the Stingray Nebula is falling rapidly.  The half-life for the decay in brightness will equal $20 / \log _2[\mathcal{F}]$ years, with $\mathcal{F}$ being for the 1996--2016 time interval.  For averages over the whole nebula, H $\alpha$ has a half-life of around 29 years, while  [\OIII] has a half-life of around 8 years.  For the regions of the Outer Shell, the half-life gets as short as 3 years.  A century from now, the H $\alpha$ flux will have dropped by an order of magnitude, while the [\OIII] flux will fall below the detection threshold.  So the duration of this entire episode of an ionized shell (which started around 1980) is only two centuries or so.  This forces the realization that this entire episode is a temporary short-duration interruption of the invisibility of the expanding PN shell.  The long-lasting blooming of the PN shell must await the central star evolving to a very high temperature millennia from now.  The 1980s ionization event can be likened to a flash bulb, briefly illuminating the PN shell.

\subsection{Balmer Decrement}

Our ratio images can map out the Balmer decrement across the surface of the PN\null.  The Balmer decrement is simply the emission-line flux ratio of H $\alpha$ to H $\beta$.  That is, from our 2016 data, we have constructed a pixel-by-pixel map of the ratio of F656N (H $\alpha$) image divided by the F487N (H $\beta$) image.  (The {\it HST\/} Instrument Handbooks give the filter transmissions and detector quantum efficiency to be similar to within a few percent for the two relevant wavelengths.)  In the standard PN case with no extinction, the decrement should closely equal 2.86.  We should only get deviations from the 2.86 due to extinction where H $\beta$ line is dimmed more that the H $\alpha$, so with extinction we get decrement $>$2.86.

For this Balmer decrement ratio image, we see a very uniform decrement across the entire nebula.  This value is close to 4.0, with an RMS scatter of 0.2.  We can calculate the equivalent $E(B-V)$ for an assumed $R_V=3.1$, as appropriate for the interstellar medium.  The observed Balmer decrement of 4.0 implies $A_V=0.8$ mag and hence $E(B-V)=0.25$ mag.  The 5 per cent RMS scatter of our Balmer decrement across the nebula corresponds to a variation in $A_V$ or $\pm$0.12 mag.  This $E(B-V)$ value is a bit high compared to other published values (see table 2 of Reindl et al.\ 2014 for values from 0.11 to 0.24).

The only exception (to the uniform Balmer decrement) is a narrow region along a 30$\degr$ arc (as seen from the central star) along the curved edge of the Outer Shell towards the south-east.  Here, the observed decrement gets up to 6.2.  This high-decrement arc closely matches an intensity feature in the nebula, so it is very unlikely to be a high-extinction feature of the ISM; rather it must be embedded in the edge of the PN shell itself.  For this high-extinction dusty area on the edge, we get $A_V=1.9$ and $E(B-V)=0.59$, which also includes the ISM component.	

So the picture we get is that of uniform coverage (to within 5 per cent) by the ISM all across the nebula (in a roughly 1 arc-second radius around the central star).  However, there is a fraction of the low-ionization edge that has a substantial extra extinction, with this component corresponding to $A_V=1.1$ mag.

\subsection{Companion Star}

A companion star appears inside the Bright Inner Ring, 0.45 arc-seconds to the northeast of the central star.  This star might be a random background star, or some distant binary partner of the central star.  The minimal separation along the line-of-sight is 700 AU (for a distance of 1.6 kpc), which translates to an orbital period of $\sim$20,000 years for a circular orbit.  With our 2016 {\it HST} images, we measure $B=17.52$, $V=16.85$, and $R=16.57$.  The companion star and the central star have identical proper motions from 1996 to 2016 to within 1 milli-arc-second per year.

For $E(B-V)$ of 0.20 mag and a distance of 1.6 kpc, the companion star has an absolute magnitude, $M_V$ of 5.36 mag and extinction corrected colors of $B-V$=0.47 and $V-R$=0.12.  This places the star well below the main sequence in the HR diagram, at a position where there are no stars.  The only way we can think to cure this problem is to place the companion star at a distance of roughly 4 kpc as an F5 main sequence star.  With the distance limits from Reindl et al. (2014), this makes the companion star a disconnected background object.

The proper motions are identical to within 1 milli-arc-second per year, whereas eight nearby stars have proper motions different by up to 10 milli-arc-seconds per year.  The probability of a random background star matching that of the central star is less than 1 per cent.  This makes a good argument that the companion star is actually in a binary with the central star.

These two arguments give different results.  One way to resolve the question could be to move the Stingray out to a distance of 4 kpc.  But this produces other problems (c.f. Reindl et al. 2014).  So we are left with no confident solution as to whether the companion star is physically associated with the Stingray.

\section{Expansion of the Nebula}

The presumed expansion velocity of the PN shell is 8.4 km/s, as taken from the [\OIII] line width (Arkhipova et al.\ 2013).  We know the distance to $\sim$50 per cent (Reindl et al.\ 2014), so we can get the physical size of the outer parts of the shell with useable accuracy.  These two quantities make the PN ejection about one millennium ago.  (That is, at 1.64 kpc, and 8.4 km/s, the PN shell will expand to 1 arc-second in 920 years, with substantial uncertainties.)  One of our purposes for {\it HST\/} imaging in 2016 was to get an expansion age, and this can be done in a distance-independent manner.  (For example, if the shell expands 1 per cent in a decade, then the shell was ejected 1000 years ago.)  As we know the star's position in the red side of the HR diagram (from the time that the PN shell was ejected) from ordinary stellar models, and we know the star's current position on the HR diagram (as well as its position in 1890), we can use our expansion age to calculate the rate of temperature change as the central star moves across the HR diagram.  If we can measure an expansion age, we can directly confront models (e.g., Miller Bertolami 2016).  And getting an expansion parallax has to be useful also.

Palen et al.\ (2002) used {\it HST\/} WFPC2 narrow-band images of three very small PNs to get measured expansions.  They used two methods:  First, they measured the gradient of the flux along radial spokes, taking the point of steepest descent to define the edge of the shell, and watched as this moved outwards.  Second, they took the old first epoch images, expanded them by a magnification factor (plus shifting and scaling), subtracted the two images, and then by eye looked for the acceptable range of magnification factors that minimized the differences between the two epochs.  We have tried both methods.

For the first method, we have generalized their method of just looking at the gradient.  In particular, we have made this into a chi-square fit that compares the radial profiles along any specified range of distances from the central star.  We could set this up for just looking at the gradients, or we can broaden it to cover the cross section of an entire arc so as to fit for the relative radial position of the peak of the arc.
	
To seek proper motion of the nebular structure, we compare a first-epoch image from 1996 to a second epoch image from 2016.  (The earlier 1992 images have the original aberrations for {\it HST\/}, so the data quality for high-resolution work is too poor.)  The image pair must be through the same narrow-band filter, so as to avoid having multiple independent lines showing different structures.  The best pair of images are through the F502N ([\OIII]) filter, as this provides the best signal.  The 1996 WFPC2 data have a nominal plate scale of 0.0455 arc-seconds/pixel.  Our 2016 WFC3/UVIS images have a plate scale (as taken from the WCS) of 0.03962 arc-second/pixel.

For the chi-square fits, we first extracted the flux from each pixel within some distance from a straight line through the centre of the central star.  For illustration here, we have chosen the north/south line (which passes perpendicularly through the edge of the arc of the Outer Shell) and the major axis of the Bright Inner Ring (which perpendicularly passes through the edges of the Bright Inner Ring and some outer arcs).  The width of the band should not be too large (so as to start blurring features) or too small (so as to get too few pixels without good resolution), so we have adopted 0.080 arc-seconds as the half-width for the band passing through the central star.  The distances along the axis are then converted from pixels to arc-seconds, through the plate scale.  The pixels values are then labelled by the distance from the central star along the direction of the line.  A plot of the pixel brightnesses as a function of distance along the line is then a measure of the radial profile of the nebula.
	
Fig. 6 shows a profile plot along the north/south axis in the lower panel, and a profile plot along the major axis of the Bright Inner Ring in the upper panel.  The green circles are for each pixel in the band inside our F502N image from 2016.  The blue diamonds are for the F502N image from 1996.  These have been scaled vertically so to roughly match as shown in the plot.  This vertical scaling is actually a free parameter in our chi-square fit, used to account for the fading of the structure from 1996--2016.  For the best accuracy, we should only look at arcs where they are fully tangential in direction.  In the lower panel for the north/south axis, we can use the steep-gradients near +1.0 and  $-1.2$ arc-seconds, as well as the local maximum (i.e., the centre of an arc structure) near +0.9 arc-seconds.  In the upper panel for the major axis, we can use the steepest gradients around $-0.9$ and +0.9 arc-seconds, as well as the local maxima around $-0.8$ and +0.8 arc-seconds.

We immediately see major problems.  First, the features are not sharply defined, so we expect poor accuracy in measuring the expansion.  Second, the shape of the profiles changes substantially from 1996--2016, so it will be difficult to even define what to measure for the expansion.  Third, and most blatant, the lower panel shows prominent peaks with apparent systematic shifts in their radial distances, but the trouble is that the 1996 profile (blue diamonds) is the one that shows the peaks farther out than the 2016 profile (green circles).  That is, it appears that the Bright Inner Ring has {\it shrunk\/} in size. A contracting PN is not physical.  So already, for these three reasons, we know that we cannot measure the expansion.
	
From the previous crude estimate that the expansion age of the Stingray PN is around a millennium, a structure at $1''$ from the central star will have a radial proper motion by $0\farcs02$ over the interval 1996--2016.  That is, we are expecting to see the radial position of the maximum gradients and peaks in the profiles to move outward by something like $0\farcs02$ or 40 per cent of a pixel.  To get a 3-sigma detection of expansion, we would need $\sim\!0\farcs007$ ($\sim$13 per cent of a pixel) accuracy in placing the structures.  A look at the profile plots in Fig. 6 suggests that it is dubious that such accuracy can be achieved.

Nevertheless, we have made a chi-square analysis to quantify the expansion.  In particular, we have compared the magnified 1996 profile against the 2016 profile over a selected range of radii, and varying the magnification factor until the difference in profiles returns a minimum chi-square.  For this, the free fit parameters are the magnification factor (which quantifies the expansion) and a brightness scaling factor (which allows for the fading).  The one-sigma uncertainty is taken to be the RMS scatter in the 1996 profile.  A substantial problem is that the profile shape changed greatly from 1996 to 2016, so the minimum chi-square is always very large, resulting in formal error bars that are unrealistically small.  For a fit from $-1.0$ to $-0.6$ arc-seconds along the profile of the major axis, the formal best-fitting magnification factor is $0.9657\pm0.0022$.  For a fit from +0.6 to +1.0 arc-seconds along the profile of the major axis, the formal best-fitting magnification factor is $0.9256\pm0.0013$.  For a fit only over the steepest gradient portion of the north/south profile, from +1.0 to +1.2 arc-seconds, the best-fitting magnification is $0.935\pm0.012$.  For a fit from +0.8 to +1.2 arc-seconds on the north/south profile (aiming to fit the peak near +1.0 arc-second), the best-fitting magnification is $0.940\pm0.010$.  These best-fittings have magnifications less than unity, which shows a {\it contraction\/} of the nebula.

These chi-square fits, as well as the gradient method of Palen et al.\ (2002), fail completely because the structures in the Stingray Nebula are all fading rapidly at greatly varying rates.  We can see what is going on in Fig. 6, where the light from the Outer Shell provides a halo brightly seen in 1996, yet invisible in 2016.  For the 1996 profile, the peaks around $\pm$0.9 arc-second have substantial additional light from a component outside, which raises the outer parts of the observed profile, shifting the apparent peak out to a larger radius.  This component has nearly completely faded away in 2016, so the preferential rising of the outer parts of the peak are absent.  This outer component only in 1996 makes for the unphysical apparent contraction of the nebula.  There is no way to model the fade rates and the profile of the two components, so no one can pull out the expansion. 

We have also tried the second method of Palen et al.\ (2002).  We have performed all the image manipulation on the 2016 images, because they allow for the best interpolation when rotating and shifting.  Only when we get a shifted/scaled/rotated 2016 image do we subtract the early-epoch image.  We have used the 1996 data for our first epoch, because the 1992 images suffer the original {\it HST\/} optical aberrations that would cripple this application. We apply the following procedures in IRAF:  (1) First, magnify by a factor of 10 to allow for better interpolation.  (2) Then rotate to the correct orientation as the 1996 images.  (3) Magnify by a factor equal to the ratio of the pixel sizes and multiplied by the near-unity magnification factor.  (4) Crop the image to the size of the cropped 1996 image.  (5) Shift so that the centres of the central star are at the same coordinates.  (6) Scale the image brightness so as to minimize the deviations from zero in the later subtracted image.  (7) Subtract the prepared 2016 image from the unchanged 1996 image.  (8) Measure the flatness of the difference image with a statistic of the RMS scatter.
	
The resulting difference images have large differences, all in a bewildering pattern.  By eye, there is no noticeable change as the extra magnification is varied over a plausible range.  The light and dark arcs on the difference images depends critically on the scaling factor (that must be included to account for the secular fading of each structure), where one scaling factor is not appropriate for most of the image (see Table 1).  Palen et al.\ only said that they did the comparison by eye, which is subjective, but the Stingray presents such a complex case as to make for no unique best case.   As we see from the profile plots above, there will always be an outer edge inside an inner black ring, and we can change the relative positions in many ways (increasing the presumed magnification, shifting the centre slightly, changing the brightness scaling), and we have no way of knowing what to look for in this complex difference-image.  The use of the RMS scatter is of no use because the value thrashes around by small changes in any of various inputs.  Palen's method works well for simple circular arcs with no fading, but this method fails for the Stingray's many arcs, rings, and nodes, all fading fast at different rates.

Simply having these many features along with a homologous expansion should be no problem for either of our two analysis methods.  The problem is that these features are rapidly fading, with each feature fading at a different rate.  In this case, the relative dimming of an outer feature will cause an apparent contraction of some inner ring, while the relative brightening of an outer feature will cause an apparent expansion on top of the real expansion.
	
So in all, we cannot get any useable expansion age or expansion parallax for the complex system of fast-fading arcs and loops and shells of the Stingray Nebula.

\section{Photometry of V839 Ara}

From the time when V839 Ara left the AGB while ejecting the PN $\sim$1000 years ago, the post-AGB motion in the HR diagram of the central star must have been horizontal to the left.  Then, V839 Ara was seen to fade slowly from 1890 to 1980 (dropping by 0.5 mag), then fade fast from 1980 to 1996 (dropping by 4 mags).  The detailed comparison of the observed position on the HR diagram versus one typical model for an LTP is shown in fig 4 of Schaefer \& Edwards (2015).  This observed evolution is very fast compared to all models.  And why is it suddenly going down fast?  We were left with possible future evolution being that the central star continues to fade fast, that the brightness evolution stops, or that the star turns around in one of those evolutionary loops and starts brightening.  For the situation just before our {\it HST\/} images in 2016, we would not be surprised by any brightness from 12 to 20 mag.

Reindl et al.\ (2017) claim from 2002--2015 {\it HST\/} spectroscopy that the central star has started getting cooler and expanding.  After considering the possibilities, they take their track in the HR diagram to be pointing to a turn in an evolutionary loop for a LTP.  

Our goal is to get the broad-band light curve of the central star {\it alone\/} after the 1980s, leading up to our {\it HST\/} magnitudes from 2016.  The light curves from the ground are unresolved, showing the combined light of the emission lines plus the central star, so some means is needed to separate out the light of V839 Ara.  This can be done with the great angular resolution of {\it HST\/} imaging.
  
Bobrowsky et al.\ (1998) used {\it HST\/} WFPC2 and a continuum filter centered at 6193~\AA\ from WFPC2 images.  They quote $V=15.4$ mag for 1996 March 8 (1996.183).

Reindl et al.\ (2014) report on an {\it HST\/} FOS spectrum taken on 1997 February 3 (1997.093), for which they give $V=14.9$ mag in their Section 4.1.  Further, Schaefer \& Edwards (2015) give $B=14.64$ and $V=14.96$ values for 1996.176 as derived from the same calibrated FOS spectrum (see fig. 2 of Reindl et al.\ 2014).

Recently, Otsuka et al.\ (2017) published UV-optical-IR spectra, for which they have taken the optical spectrum from their FEROS spectrum on the ESO 2.2m telescope on 2006 April 16 (2006.290).   They pulled out the central star alone by subtracting the emission lines and the nebular continuum.  They then integrated over the filter transmission curves of the {\it BVRI\/} bands.  For example, they quote a de-reddened $V=14.51\pm0.17$ in their table 9.  With $E(B-V)=0.11$ and $R_V=3.1$ for their de-reddening, we have $A_V=0.341$ and an observed $V=14.178$.  The calculated error bars range from $\pm$0.12 mag in $B$ to $\pm$0.44 mag in $I$.

Here, we add the $BVr'$ broad-band photometry from our WFC3/UVIS images on 2016 February 21 (2016.058) as well as $V$-band photometry from the WFPC2 images on 2000 March 2 (2000.169).  We have performed the usual {\it HST\/} `absolute photometry,' where we count the flux in photoelectrons inside some standard-sized circular photometry aperture and use the calibrated zero point to convert to a standard magnitude.  Here, we use the Vega magnitude system.  The dominant uncertainty is due to not knowing the effective background flux appropriate for the sky under the star in the photometry aperture.  The sky background outside the Stingray is close to zero.  The central region of the Stingray, inside the Bright Inner Ring, apparently has a broad and flat minimum.  This minimum is much higher than the sky value outside the Stingray, and its flux is a significant fraction of the total flux inside the photometry aperture.  Unfortunately, there are few pixels between the tail of the PSF of the central star and the inner edge of the Bright Inner Ring.  Indeed, there is no assurance that the surrounding pixels are really at a minimum appropriate for the area inside the central star's photometry aperture.  So all we can do is take the flux from the surrounding pixels (in particular, the pixels that are with a minimal flux around the star) and hope that this is about right.  We have no way of getting a formal error bar.  By taking some plausible range for the uncertainty of the applicable background, the resultant uncertainty in the magnitude is around 0.1 mag.  A specific problem is that the image of V839 Ara in the 180-second $V$-band image from the year 2000 does not have the usual PSF profile, rather appearing as a large flat-topped shape, so something went wrong here, perhaps a cosmic ray, so this image was not used for photometry.  For the 2016 images taken back-to-back, we report the average magnitudes.

Our summary of all the broadband $BVRI$ magnitudes {\it for the central star alone\/} are presented in Table 2.  From 1996 to 2016, we do not see any monotonic progression of magnitudes, nor even a simple turn-around associated with an LTP loop.  Rather, it appears that V839 Ara is roughly holding constant in brightness, with some erratic variability superposed.  This variability is nominally much larger than all measurement errors, but we are concerned about the cross calibration of methods and instruments from many different sources, all in a complex situation with the embedding nebulosity.  It is tempting to try to ignore some selected fraction of the light curve so as to match some preconceived evolution, but there is no evidential basis for throwing out any of the magnitudes in Table 2, we would have to throw out many magnitudes to fit any smooth curve, and such a procedure could only be wishful thinking.  In all, we think that this variability is real.  But we do not understand the observed fast variability, nor do we even have speculative ideas as to the cause for these fast ups-and-downs.
	
With our photometry, we have not seen any decisive movement in the light curve.  In particular, with Reindl et al.\ (2017) demonstrating that V839 Ara has switched its progression in 2002, starting to cool and expand.  Our light curve does show a cooling, with the $B-V$ changing from $-0.32$ in 1997, to $-0.27$ in 2006, and $-0.18$ in 2016.  But we do not show the luminosity increase.  We do not understand this lack of brightening.

\section{How to Hide the 1980's Ejecta?}

We have the strong prediction that the outgoing 1980s ejecta must form a shell well away from the central star and already be ramming into the inner side of the old PN shell.  That is, for a velocity of 1800 km s$^{-1}$ and a distance of 1.6 kpc, the ejecta will reach 1 arc-second from the central star years before our 2016 images.  The ejecta should be blatantly visible by means of either seeing the gas in the expanding shell, or by the emission arising from the shocks.  But we looked in many ways and see no evidence of the 1980s ejecta.  This now becomes a problem for how to resolve this paradox.

One possible explanation is that the 1980s ejecta had already suffused through the nebula by the time of the 1996 images.  At 1800 km s$^{-1}$ and 1.6 kpc, the ejecta would pass the Bright Inner Ring (with a semi-major axis of 0.75 arc-seconds) after 3.1 years.  If the ejecta started out in the early 1980s, then the collisional shock should already have lit up the PN long before the first {\it HST\/} images in 1992.  With this, we would not have seen any changes in the {\it HST\/} images from the impact starting up, nor would we see any sudden brightening in the 1994--2015  $V$-band light curve, and the shell of ejected material would have had time to suffuse through the entire nebula with no isolated trace.  Nevertheless, this early-impact idea might fail if the ejecta sweep up a thin shell of high-density gases (which is not seen), or if the shocked region breaks up into clumps due to Rayleigh-Taylor instabilities (which is not seen).  Still, it is possible that the clumps are not resolved, that the only substantial close-in gas was in the Bright Inner Ring which is now lit up, and the various outer arcs are the shocked shells away from the plane.  So it is unclear whether this explanation can be made to work.

Another possible explanation is that the total mass of the 1980s ejecta might be sufficiently small so that the ejecta is not directly visible.  The outflow apparently started around 1980, continuing at a high rate through to the early 1990s.  The mass-loss rate was measured at $3\times10^{-8}\, $M$_{\odot}\,\rm yr^{-1}$ in 1988 (Feibelman 1995), was measured at near 10$^{-9}$ M$_{\odot}$/year in 1988, 1992, and 1993, tapering off to 10$^{-10}$ M$_{\odot}$/year in 1994, 1995, and1996, while falling further to below 10$^{-11.3}$ M$_{\odot}$/year in 2002 and 2006 (Reindl et al.\ 2014).  If we presume an equivalent mass loss at the 1988 rate for a decade, then the mass ejected is around $3\times10^{-7} $ M$_{\odot}$.  This total mass is much smaller than the $\sim$$10^{-5}$ M$_{\odot}$ shells ejected by classical novae (Yaron et al. 2005).  With this experience from novae, we would not expect to see the outgoing ejecta as a shell of gas.  Nevertheless, this is a weak precedent because almost all novae do not include the collision of high-velocity ejecta with a pre-existing shell.  The 1980s ejecta are also many orders of magnitude less massive than the shell of the PN, which is $0.038 $M$_{\odot}$ for both the neutral and ionized gas (Otsuka et al.\ 2017).  The kinetic energy of the ejecta will only be sufficient for one outgoing hydrogen ion to ionize close to a thousand hydrogen atoms in the PN, so our estimate of the outgoing shell mass will allow for the creation of only $\sim\!3\times10^{-4} $M$_{\odot}$ of ionized gas.  It is unclear whether the collisional ionization from the 1980's ejecta can cause the PN to light up with emission lines as seen.  Detailed calculations are needed, and such should also consider the case where the initial ejections from 1980--1988 are substantially higher than first seen in 1988.  Still, this offers a possible way to hide the 1980's ejecta as having too small mass to be directly imaged, even though the kinetic energy of the ejecta might appear as a result of collisional ionization by the shocks produced by the ejecta.

\section{Nature of the Ionizing Event of the 1980s}

Previously, the nature of the mechanism for the ionization event has either been completely ignored or been declared to be a fundamental unanswered problem (Schaefer \& Edwards 2015).  The obvious mechanism to ionize the Stingray is that the central star produced a high luminosity of far-ultraviolet light during the 1980s, just as ordinary PN are being ionized.  The only other mechanism for the sudden ionization of the PN shell is that some outgoing shell of mass ejecta has collisionally ionized the pre-existing PN shell.  So the mechanism is either radiative ionization or collisional ionization.

PN shells are usually ionized by the UV radiation emitted by the central star, as the star heats up to high temperatures.  This radiation, with wavelengths below the Lyman limit, is the ordinary blackbody emission from the hot stellar surface.  In principle, some additional source of ionizing radiation might ionize the PN, but we have not even heard of speculation on such possibilities.  What is needed is for the ionizing flux to turn-on around 1980, reach some high peak UV flux in the 1980s sufficient to ionize the entire PN shell, then to fall-off rapidly at least after 1993.  However, the mechanism of radiative ionization is unlikely; for four strong reasons:  (1) First, the $B$-band light curve was well sampled from 1969 to 1989 with the Harvard plates, and no increase in the $B$-band flux was seen (Schaefer \& Edwards 2015), whereas any plausible spectrum for the ionizing flux should also produce a large luminosity of extra light in the blue.  This rules out ionizing radiation from any source, unless its duration was less than a few months and it flashed around the time of solar conjunction.  (2) Second, the surface temperature of the central star was  measured to be around 28,000 K in 1971 and 38,000 K in 1988 (Schaefer \& Edwards 2015; Reindl et al.\ 2014), and these temperatures are too low to create a useful luminosity of ionizing flux, much less for full ionization on the time scale of a year or so.  In particular, we can calculate the flux for light below the Lyman limit throughout the last century based on the temperatures and luminosities as given in table 3 of Schaefer \& Edwards (2015).  This ionizing luminosity (in units of the solar luminosity) is 1200 in 1889, 430 in 1980, 160 in 1988, 300 in 1996, 120 in 2002, and 80 in 2006.  The point is that there is no sudden surge of ionizing radiation in the 1980s.  (3) Third, the mass loss was still going in 1988, when \IUE\/ showed that there was no substantial extra ionizing radiation in the far UV (Feibelman 1995).  (4) Fourth, the central star's temperature increased steadily up to 60,000 K in the year 2002 (Reindl et al.\ 2014), but the nebula is still fading, proving that the starlight is not the source of the ionization of the Stingray.  So the obvious mechanism (a sudden high luminosity of far-ultraviolet radiation in the 1980s) does not work.  That is, we have observed in various ways that any speculated ionizing flux did not exist.

The only other possible explanation for the sudden ionization of the nebula around 1980 is simply to say that the entire 1980s ionization event is the result of earlier ejecta ramming into the PN shell, making for collisional ionization through shocks.  With this possibility, we get one solution to two of the fundamental questions; `How were the 1980s ejecta hidden?' and 'What is the mechanism for the 1980s ionization?'.  In this scenario, the 1800 km s$^{-1}$ outflow started around 1980, and within three years started shocking the pre-existing PN shell, with the shocks ionizing the Stingray.  Whatever mechanism on the star that caused the outflow also made for the fading of the star.  This outflow continued until the end of the 1993 or so, with the ejecta suffusing the nebula and the shocks spreading throughout the Stingray.  We could say that the ejecta were not hidden, because it was blaring forth as the incredibly bright emission lines that caught the attention of the world.  Alternatively, we could consider that merely measuring the effects of the ejecta is not {\it seeing\/} the ejecta, in which case we would answer the question by saying that the ejecta were hidden because the first {\it HST\/} images came long after the outgoing gases had become invisible as a separate entity.  This scenario also explains why the mass-ejection and the ionization occurred simultaneously.

The sudden ionization of the Stingray around 1980 could either be caused by a sudden turn-on of some luminous UV light source, or by some set of impact shocks that collisionally ionize the gas.  Well, we have just shown that the UV light source did not exist, so the only possibility left to ionize the Stingray is collisional ionization.

This scenario (the collisional ionization of the Stingray from the 1980s outflow) leaves unanswered the questions as to the mechanism for pushing the outflow and {\it why\/} the outflow started.  (The alternative scenario of the sudden turn-on of a high-luminosity source of far-UV light has the identical problem in that the originating mechanism is unknown.)  We wonder whether there might be tests for this scenario possible from plasma diagnostics, where collisional ionization from shocks might produce a different early spectrum than would ionization from radiation.  Certainly, before it can be accepted, this scenario must pass tests of possibility as based on physical models.

In the end, we think that the ionization of the PN shell starting around 1980 is due to the shocks made by the collisions of the observed ejecta hitting the PN gas.  Our reason is that the only alternative explanation is already rejected for four observational refutations.  Further, by Occam's Razor, the collisional-excitation idea requires no new input hypotheses (because the outgoing ejecta was observed and is adequate), whereas the alternative hypothesis requires the presumption of an invisible light source from an unprecedented new mechanism of mysterious origin.

\section*{Data Availability}

The data in this paper are available in the public archives of the {it Hubble Space Telescope\/} at \url{http://archive.stsci.edu/}, or through the Mikulski Archive for Space Telescopes at \url{https://mast.stsci.edu/}.

\section*{Acknowledgements}

Support for programme number GO-14126 was provided by NASA
through grants from the Space Telescope Science Institute, which is
operated by the Association of Universities for Research in Astronomy,
Inc., under NASA contract NAS5-26555.

We thank Zoltan Levay for preparing Fig. 1 and 2 from the \HST\/ images.


{}


\begin{figure*}
	\includegraphics[width=6.0in,angle=-90]{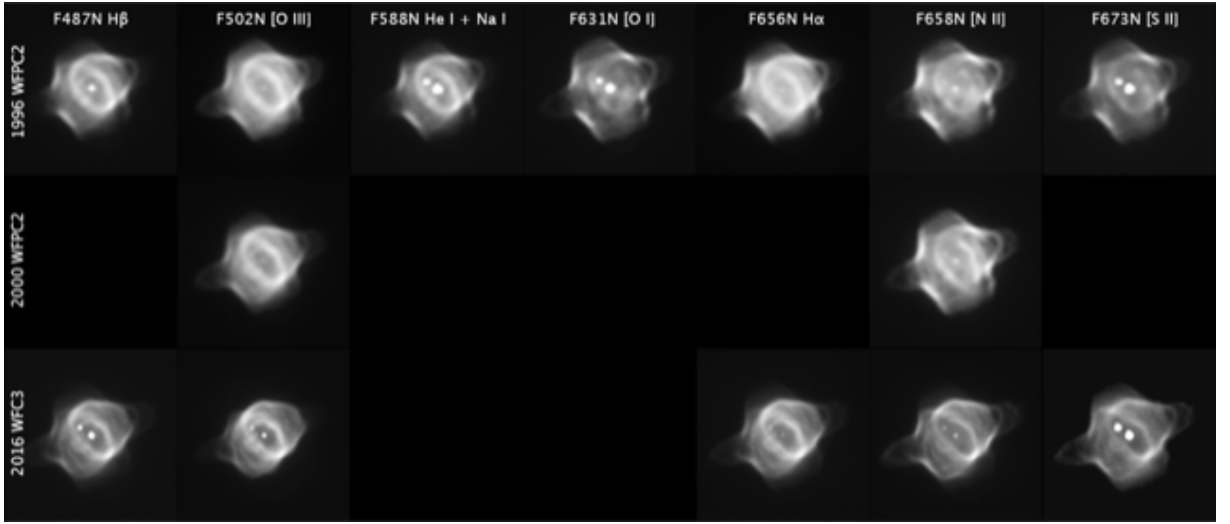}
    \caption{The {\it HST\/} narrow-band images.  This montage shows thumbnails of most of the {\it HST\/} images, all with a uniform scale and orientation (north to the top, east to the left).  They are organized with the top row for the 1996 WFPC2 images, the middle row for the 2000 WFPC2 images, and the bottom row for our 2016 WFC3 images.  The 1992 WFPC images and the $BVr'$ broad-band images are not shown.  Each column displays the images with various filters, as labelled at the top of each column.  The grey scale is chosen so that maximal detail is visible in the nebula.  Such a choice hides that the nebula is fast fading.  Further, the grey scale choices can be misleading in terms of the brightness of the central star.  The relative brightness between the nebula and the stars changes in proportion to the filter width.  A wide filter lets in a lot of starlight, while a narrow filter lets in the same amount of nebular light from the very narrow lines, but lets in only a small amount of star light.  So the central star is very prominent in the [\SII] images because the F673N filter is the broadest (at 77 \AA).  The smallest-width filters are for H $\alpha$ (at 14 \AA) and [\NII] (at 20 \AA), and the central star is very dim.  The other filters are all variously from 45 \AA\ to 54 \AA, and the central star is at middle prominence.  Still, the adopted grey scales are the best way to see the available images and what they show.  For example, we see that the outer regions are easily visible in 1996, but have greatly faded (relative to the inner features) by 2016, indicating a greatly faster recombination time-scale.  We can see that no feature brightens after 1996, with the same result for no-brightening after 1992, as evidence for the 1980s wind {\it not\/} ramming into the pre-existing PN after 1996.  For each epoch, the Bright Inner Ring is particularly prominent (compared to the outer features) in the light of [\OIII], \HeI, and the hydrogen lines, and relatively faint in the light of [\SII], [\NII], and [\OI].}
\end{figure*}

\begin{figure*}
	\includegraphics[width=6.0in,angle=-90]{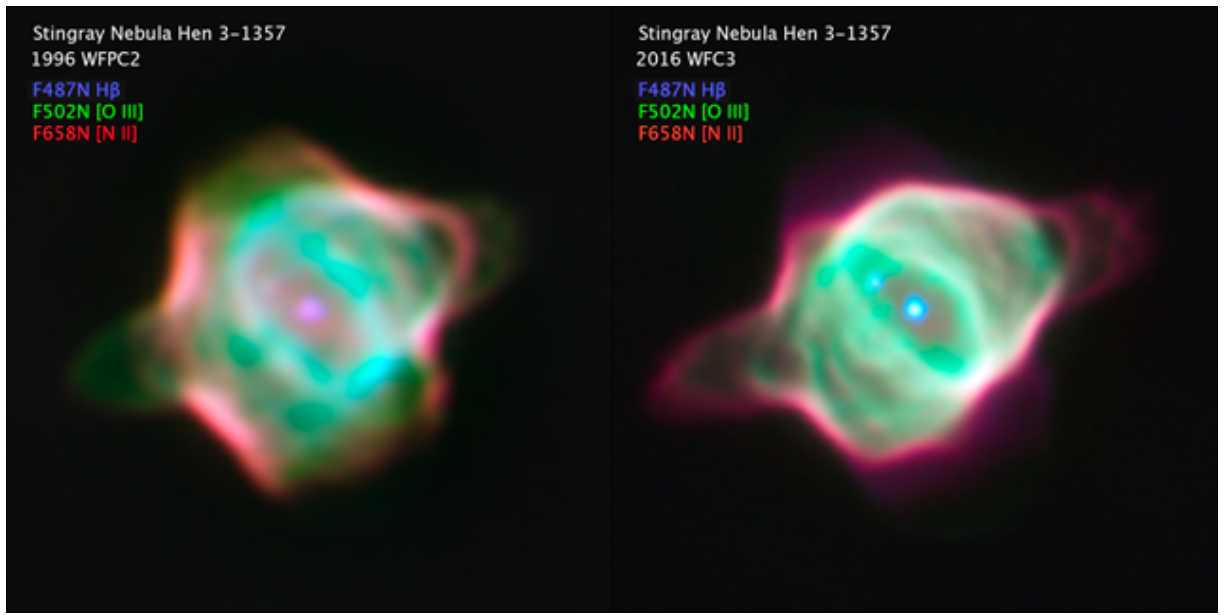}
    \caption{The combined narrow-band images for 1996 and 2016.  In this false-colour representation, the blue image plane is constructed from the F487N H $\beta$ image, the green plane is constructed from the F502N [\OIII] image, and the red plane is constructed from the F658N [\NII] image.}
\end{figure*}

\begin{figure*}
	\includegraphics[width=\textwidth]{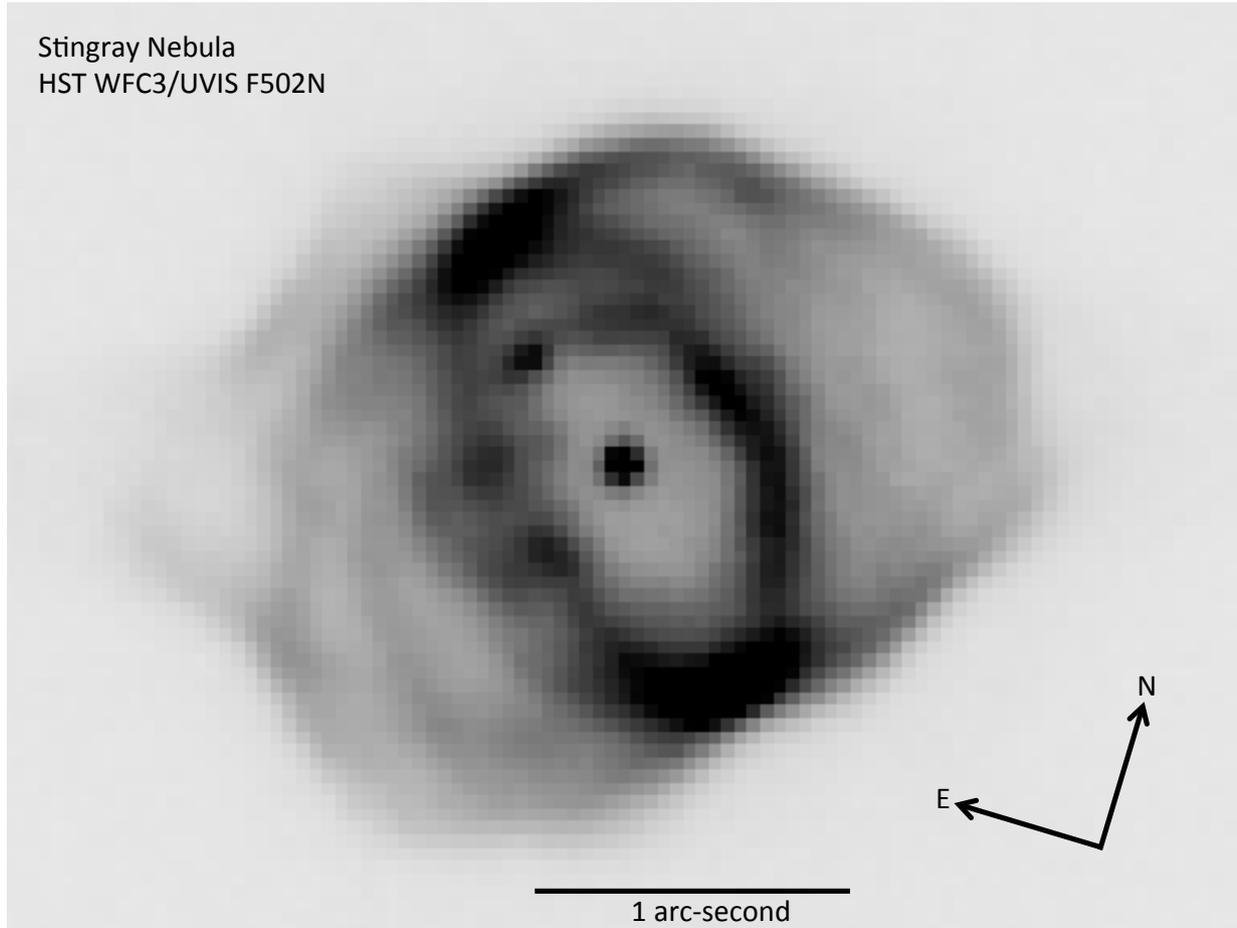}
    \caption{The Stingray in [\OIII] light.  This 60-second F502N image with the WFC3/UVIS (with 0.03962" per pixel) on 21 February 2016 shows a complicated structure of shells, arcs, a central star, plus a companion star.  A gazetteer for feature names is presented in Bobrowsky et al.\ (1998).  The most prominent feature is an ellipse, called the `Bright Inner Ring', stretching from the north-east to the southwest.  Inside this Bright Inner Ring, we only see the central star, a nearby companion star, and a `Spur' passing across the companion, all superposed on an apparently flat central background.  One point of this image is that we do not see any shell of expanding ejecta inside the Bright Inner Ring (or elsewhere).  Another point from this image is that the Outer Collimated Flow (outside the north-east edge and outside the ESE edge) are invisible, or more specifically, much fainter than inner structures as compared to the comparable 1996 image, showing that the various arcs in the nebulosity are fading at different rates. }
\end{figure*}

\begin{figure*}
	\includegraphics[width=\textwidth]{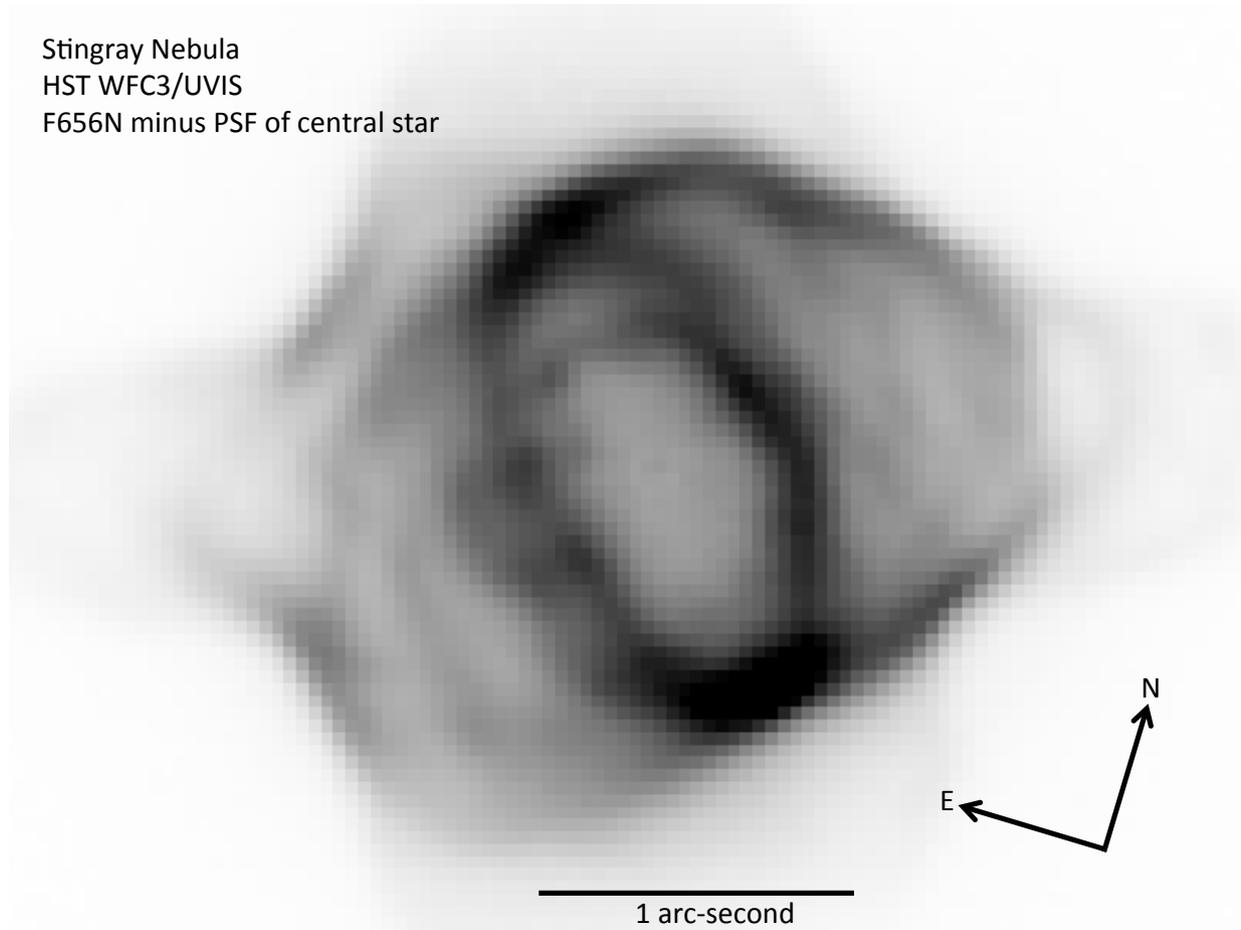}
    \caption{H $\alpha$ image with the PSF for the central star subtracted.  The base image is the 382-second F656N image from 21 February 2016 with the {\it HST\/} WFC3/UVIS.  The PSF was taken to be the scaled and shifted F625W ($r'$) image, for which the central star dominates so that very little nebular flux is subtracted.  The result should show any shell from the 1980s ejecta that underlies the central star light.  The lack of any excess flux in the central region (inside the Bright Inner Ring) shows that the narrow emission lines from the shock is producing less than 45 parts-per-million of the H $\alpha$ flux from the entire nebula.  In comparison with Fig. 3, we see no qualitative differences, so both H $\alpha$ and [\OIII] are showing the same structure.}
\end{figure*}

\begin{figure*}
	\includegraphics[width=\textwidth]{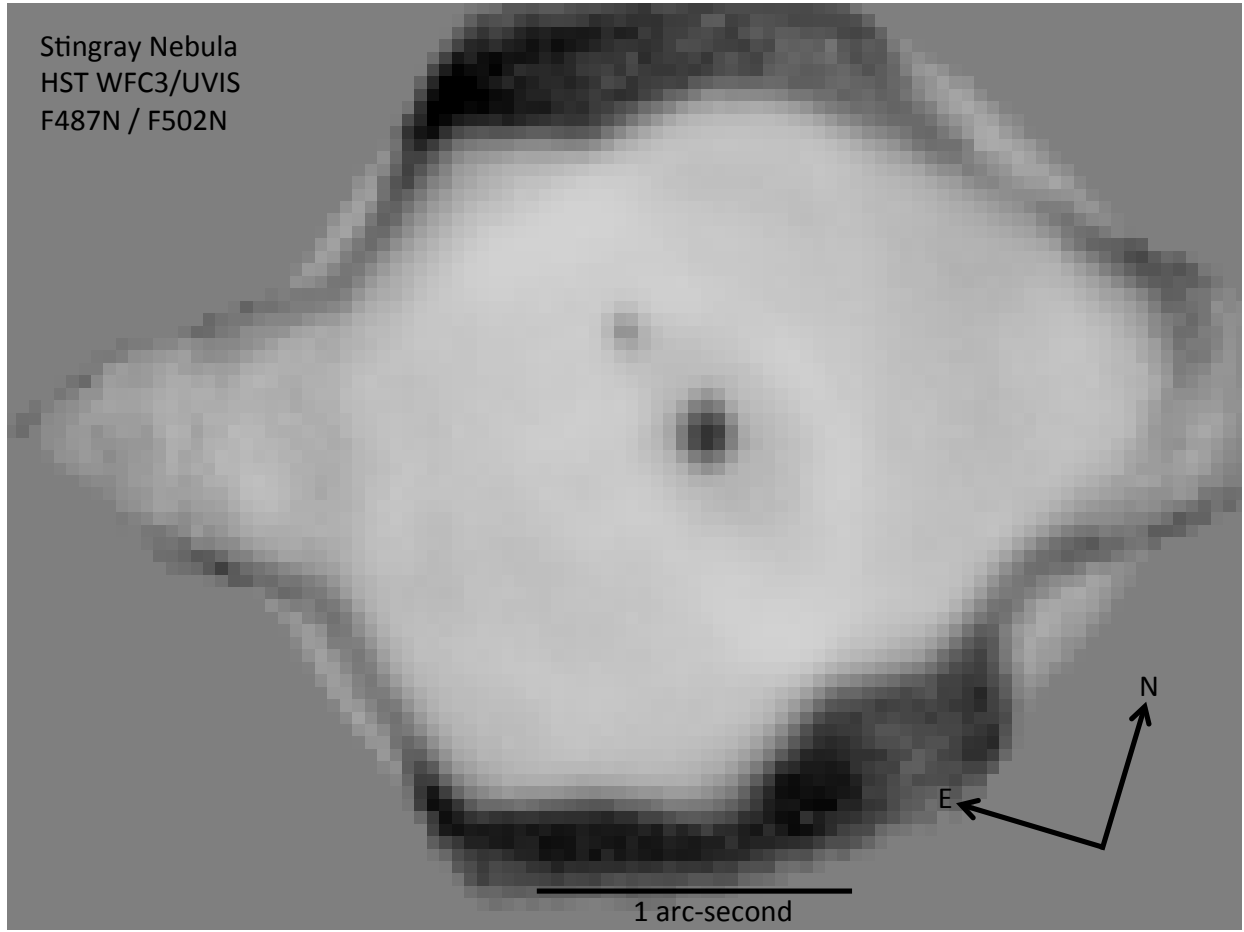}
    \caption{Ratio image of F487N (H $\beta$) divided by F502N ([\OIII]).  With the grey scale, the dark or black regions have relatively high H $\beta$ emission and relatively low [\OIII] emission.  The entire interior of the Stingray nebula has a uniform ratio, represented by a relatively bright or white region, with the [\OIII] dominating over the H $\beta$.  This shows that the Stingray has ionization stratification, where the emission lines with relatively high ionization potential (like [\OIII] dominate in regions close to the central star, while emission lines with relatively low ionization potential (like H $\beta$) dominate in regions far from the central star.  Most of the outer black areas are in the Outer Shell, where the nebulosity is greatly fainter than the easily visible interior.  That is, the low ionization regions are outside of the basic structures visible in Fig. 1 to 4.  What is going on is that the Outer Shell regions have the [\OIII] flux fading by up to a factor of 99$\times$ from 1996-2016, while the hydrogen emission faded by only a factor of $\sim$2$\times$, so for some unknown reason the [\OIII] light from the Outer Shell has largely vanished.  The F658N ([\NII] and F656N (H $\alpha$) images produce similar ratio images when compared to the F502N ([\OIII]) image, again illustrating ionization stratification.}
\end{figure*}

\begin{figure*}
	\includegraphics[width=\textwidth]{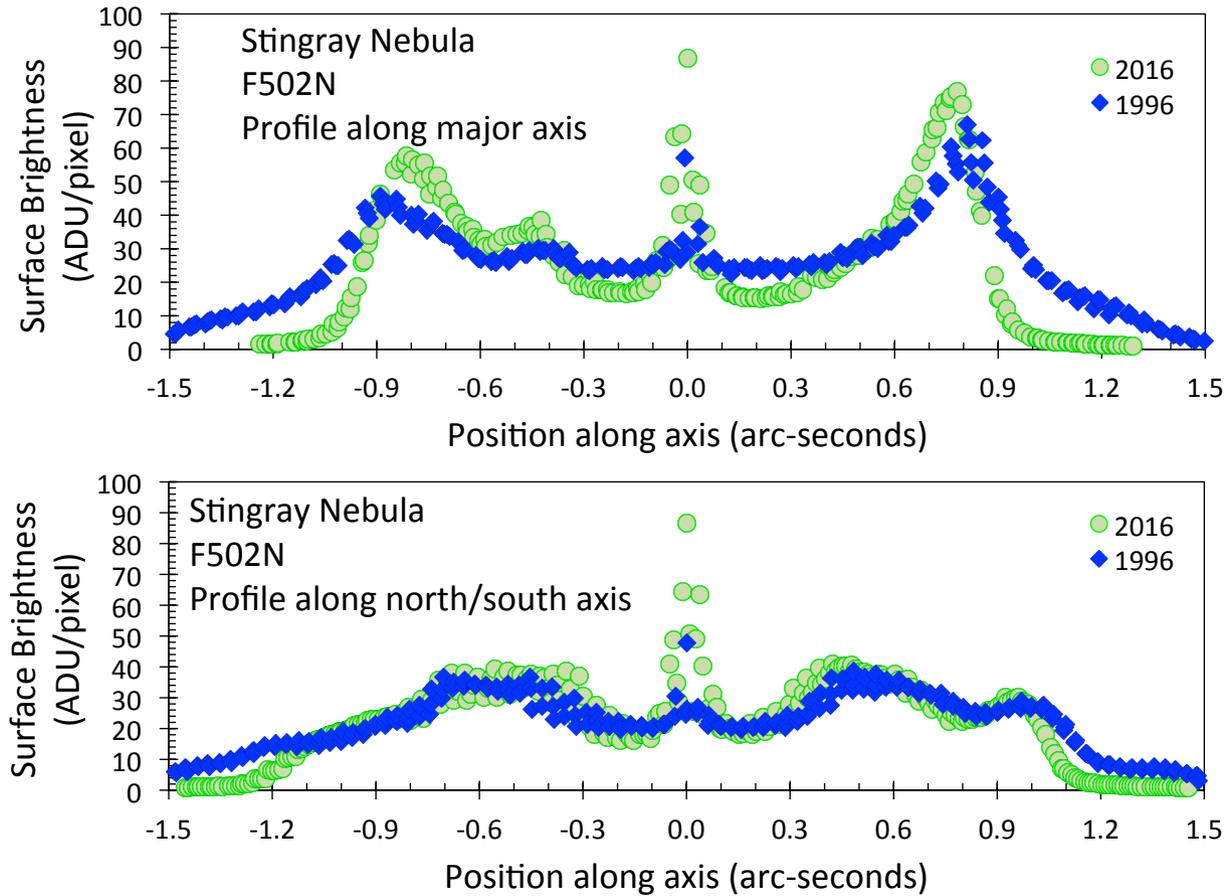}
    \caption{Profiles along two axes from 1996 and 2016.  These profiles show the surface brightness along a cut through the Stingray along an axis passing through the central star.  The 1996 profiles have been arbitrarily scaled vertically so as to allow overlap for easy comparison.  In principle, the 1996 profile (blue diamonds) should be expanded horizontally as part of a homologous expansion to match the 2016 profile (green circles).  In practice, these profiles cannot be used effectively to measure the expansion of the Stingray.   Part of the reason is that the features are too broad to allow for any accurate measure.  Another part of the reason is that the various components are fading rapidly at different rates.  With this, the Outer Shell regions, roughly outside of 1.1 arc-seconds from the central star, were prominent in 1996, but largely invisible in 2016.  This changes the shape of the profiles, and raises the outer portions so as to differently shift the apparent peaks.  This problem is starkly seen in the upper panel, where the 1996 profile peaks around $\pm$0.9" are {\it outside} the 2016 peaks, making for an apparent unphysical {\it contraction\/} of the nebula.  We conclude that we cannot make a measure of the Stingray's expansion with any useable accuracy.}
\end{figure*}

\begin{table*}
	\centering
	\caption{Fading Rates 1996--2016}
	\begin{tabular}{llllll} 
		\hline
 &  [O III] & 1996--2016 & 2000--2016 & H $\alpha$ & 1996--2016 \\
Point in Stingray Nebula & F(2016) & $\mathcal{F}$ & $\mathcal{F}$ & F(2016) & $\mathcal{F}$ \\
		\hline
Middle of Bright Inner Ring, on major axis, just SW of Central Star	&	15.5	&	9.7	&	6.7	&	15.6	&	2.0	\\
On Spur, inside Bright Inner Ring, offset from the Companion Star	&	39.5	&	4.6	&	4.1	&	27.8	&	1.3	\\
Bright Inner Ring, on major axis to the SW	&	77.0	&	5.0	&	4.5	&	57.2	&	1.0	\\
Bright Inner Ring, on major axis to the NE	&	58.8	&	4.8	&	4.2	&	42.0	&	1.0	\\
Bright Inner Ring, on minor axis to the NW	&	52.4	&	6.4	&	5.1	&	38.6	&	1.3	\\
Bright Inner Ring, on minor axis to the SE	&	43.7	&	6.7	&	5.0	&	32.8	&	1.5	\\
Faintest part between Bright Inner Ring and Outer Shell to SE	&	15.2	&	7.8	&	5.9	&	14.3	&	1.5	\\
Interior of Bubble, outside the Bright Inner Ring towards NW	&	16.2	&	10.6	&	7.5	&	16.3	&	1.7	\\
Interior of Outer Shell, outside Bubble, towards the WNW	&	12.5	&	8.7	&	6.7	&	12.6	&	2.0	\\
Just inside arc at edge of Outer Shell towards the SE	&	11.5	&	10.4	&	6.8	&	12.3	&	1.9	\\
Bright edge of Outer Shell to SE (with high {\it in situ} dust extinction)	&	18.1	&	9.2	&	6.3	&	21.3	&	1.8	\\
Bright edge of Outer Shell to ENE	&	11.9	&	13.5	&	8.9	&	16.4	&	2.3	\\
Arc at edge of Outer Shell towards W	&	29.1	&	6.9	&	5.0	&	29.2	&	1.6	\\
Arc at edge of Outer Shell towards N	&	32.2	&	6.2	&	5.7	&	30.6	&	1.5	\\
Just outside Bright Inner Ring along major axis to NE	&	0.7	&	99	&	59	&	3.6	&	2.2	\\
Just outside Bright Inner Ring along major axis to SW	&	1.0	&	78	&	39	&	3.5	&	1.6	\\
Between Inner Collimated Outflow and Outer Shell on WNW side	&	2.6	&	13.5	&	9.3	&	3.3	&	1.8	\\
Between Inner Collimated Outflow and Outer Shell on ESE side	&	3.7	&	8.5	&	6.8	&	4.7	&	1.5	\\
Between Inner and Outer Collimated Outflows on WNW side	&	1.2	&	17.0	&	10.0	&	1.5	&	1.7	\\
Between Inner and Outer Collimated Outflows on ESE side	&	1.3	&	11.4	&	7.8	&	1.4	&	1.0	\\
		\hline
	\end{tabular}
\end{table*}

\begin{table}
	\centering
	\caption{Broad-band Photometry of the Central Star Alone}
	\begin{tabular}{llllll} 
		\hline
		Year & $B$ (mag) & $V$ (mag) & $R$ (mag) & $I$ (mag)   & Source\\
		\hline
1889	&	10.30	&	$\ldots$	&	$\ldots$	&	$\ldots$	&	Harvard plates	\\
1980	&	10.76	&	$\ldots$	&	$\ldots$	&	$\ldots$	&	Harvard plates	\\
1988	&	$>$12.6	&	$\ldots$	&	$\ldots$	&	$\ldots$	&	Harvard plates	\\
1996.183	&	$\ldots$	&	15.4	&	$\ldots$	&	$\ldots$	&	{\it HST\/} WFPC2	\\
1997.093	&	14.64	&	14.96	&	$\ldots$	&	$\ldots$	&	{\it HST\/} FOS	\\
2000.169	&	$\ldots$	&	15.1	&	$\ldots$	&	$\ldots$	&	{\it HST\/} WFPC2	\\
2006.290	&	13.91	&	14.18	&	14.26	&	14.55	&	FEROS on ESO 2.2-m	\\
2016.058	&	15.32	&	15.50	&	15.79	&	$\ldots$	&	{\it HST\/} WFC3/UVIS	\\
		\hline
	\end{tabular}
\end{table}

\bsp	
\label{lastpage}
\end{document}